\documentclass[twocolumn,prl,doubespace,floatfix,superscriptaddress,nofootinbib]{revtex4} 

\usepackage{graphicx}
\usepackage{rotating}
\usepackage[usenames,dvipsnames]{color}
\usepackage{multirow}
\usepackage{amsmath,amsfonts,amssymb}

\newcommand{\remove}[1]{}

\begin{document}

\title{On the frequency and severity of interstate wars}
\author{Aaron Clauset}
\email{aaron.clauset@colorado.edu}
\affiliation{Department of Computer Science, University of Colorado, Boulder, CO 80309, USA} 
\affiliation{BioFrontiers Institute, University of Colorado, Boulder, CO 80303, USA} 
\affiliation{Santa Fe Institute, Santa Fe, NM 87501, USA} 

\begin{abstract}
Lewis Fry Richardson argued that the frequency and severity of deadly conflicts of all kinds, from homicides to interstate wars and everything in between, followed universal statistical patterns:\ their frequency followed a simple Poisson arrival process and their severity followed a simple power-law distribution. Although his methods and data in the mid-20th century were neither rigorous nor comprehensive, his insights about violent conflicts have endured. In this chapter, using modern statistical methods and data, we show that Richardson's original claims appear largely correct, with a few caveats. These facts place important constraints on our understanding of the underlying mechanisms that produce individual wars and periods of peace, and shed light on the persistent debate about trends in conflict.
\end{abstract}

\maketitle

\begin{figure*}[t]
\begin{center}
\includegraphics[scale=0.405]{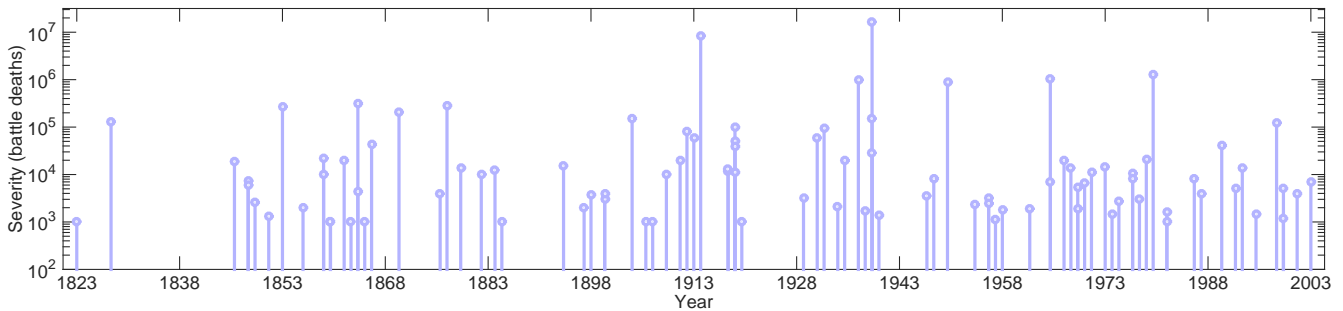}
\end{center}
\vspace{-3mm}
\caption{The Correlates of War (CoW) interstate war data~\cite{sarkees:wayman:2010} as a conflict time series, showing both severity (battle deaths) and onset year for the 95 conflicts in the period 1823--2003. The absolute sizes of wars range from 1000, a minimum by definition, to 16,634,907, the recorded battle deaths of the Second World War. Delays between consecutive war onsets range from 0 to 18 years, and average 1.91 years. Most wars (79\%) ended no more than 2 years after their onset. Originally published in Ref.~\cite{clauset:2018}. \label{fig:cow:data} } 
\end{figure*}

\noindent Lewis Fry Richardson (1881--1953) stands as one of the founding fathers of the modern field of complexity science~\cite{mitchell:2011}, which aims to understand both how complexity arises from the interaction of simple rules and how structure emerges from the chaos of contingency. One of his most celebrated works was his analysis of the frequency and severity of interstate wars and other deadly conflicts~\cite{richardson:1944,richardson:1948,richardson:1960}. Richardson also played critical roles in two other major pieces of complexity science, which continue to inform scientific efforts to understand systems as varied as developmental biology, the formation of galaxies, and the collective behavior of humans in its many forms.

The first of these arose in his work on the ``coastline paradox,'' which is captured by a deceptively simple question:\ how long is the British coastline? Richardson showed that the length of a coastline depends, paradoxically, on the length of the ruler used to measure it---the shorter the ruler, the longer the coastline's total length. This effect, now called the Richardson effect, paved the way for Benoit Mandelbrot's celebrated work on fractal geometry~\cite{mandelbrot:1967}, which has informed numerous studies of complex social, biological, and technological systems~\cite{mitchell:2011}. Richardson's insight also foreshadowed his discovery of a ``scale-free'' pattern in the statistics of wars.

The second arose from Richardson's pioneering work in meteorology, which was his primary focus for many decades. Much of his work here aimed to develop the mathematics of weather forecasting, and in recognition of those contributions, a dimensionless quantity related to buoyancy and shear flows in turbulent systems is called the Richardson number. Richardson also pioneered the use of numerical approaches to forecast the weather~\cite{richardson:1922}, despite the fact that sufficient computing power to make useful weather predictions would not be developed until several decades later. In this way, Richardson very nearly discovered, almost a half century earlier, the same mathematical chaos lurking in the equations of turbulence that Edward Norton Lorenz would later make world famous~\cite{lorenz:1963}. Richardson's work on weather forecasting also foreshadowed his interest in the long-term statistics of wars.

These intellectual threads came together in Richardson's foundational studies of violent conflict, in which he argued that the frequency and severity of deadly ``quarrels'' of all kinds, from small-scale events like homicides to large-scale events like interstate wars, followed universal statistical patterns~\cite{richardson:1960}. Although little attention is now paid to his claims about small-scale events like homicides, Richardson's ideas about larger events have become central to the study of political conflict, including civil unrest~\cite{biggs:2005}, terrorism~\cite{clauset:etal:2007}, insurgency~\cite{bohorquez:etal:2009}, civil wars~\cite{cederman:2003,lacina:2006}, and interstate wars~\cite{cederman:2003,cederman:etal:2011,pinker:2011,harrison:wolf:2012}. In this chapter, we focus on Richardson's ideas about the statistics of interstate wars.

Richardson's original analysis only covered interstate wars from 1820--1945~\cite{richardson:1948}. On the basis of these events, he made two claims about their statistical pattern. First, he argued that the sizes, or ``severities,'' of these wars followed a precise pattern, called a power-law distribution, in which the probability that a war kills $x$ people is $\Pr(x) \propto x^{-\alpha}$, for all $x\geq x_{\min}>0$, and where \mbox{$\alpha>1$} is called the ``scaling'' parameter. Second, he argued that the timing of wars followed a simple Poisson process, implying both a constant annual probability for a new war and a simple geometric distribution for years between wars~\cite{richardson:1944}.

Although his statistical methods were not rigorous by modern standards and his data were far less comprehensive, these patterns---a power-law distribution for war sizes and a Poisson process for their onsets---represent a simple and testable model for the statistics of interstate wars worldwide. Crucially, Richardson's model is ``stationary,'' meaning that the rules of generating new wars do not change over time.

If the empirical statistics of interstate wars really do follow the simple patterns claimed by Richardson, it would indicate strong constraints on the long-term dynamics of the underlying social and political mechanisms that generate wars and periods of peace~\cite{ray:1998,ward:etal:2007,leeds:2003,jackson:nei:2015,alesina:spolaore:1997,jackson:nei:2015}. A long-running debate within the study of conflict has focused on whether or not such conflicts are characterized by genuine trends. (See Ref.~\cite{gleditsch:clauset:2018} for a recent review of this debate.)

If the underlying mechanisms that produce wars are stationary, then any ``trend'' is inherently illusory. However, deciding whether trends exist has proved difficult to resolve, in part because there are multiple ways to answer this question, depending on what type of conflict is chosen, what variable is analyzed, and how the notion of change is formalized. Different choices can lead to opposite conclusions about the existence or direction of change in the statistics of conflict~\cite{payne:2004,harrison:wolf:2012,braumoeller:2013,cirillo:taleb:2016,gleditsch:clauset:2018,clauset:2018}.

Here, we consider a more straightforward question:\ given modern statistical tools and interstate war data, do Richardson's claims about statistical patterns hold up, and if so, what does that imply about the long peace of the post-war period? For this investigation, we apply state-of-the-art methods~\cite{clauset:etal:2009,clauset:woodard:2013} to the set of interstate wars 1823--2003 given in the Correlates of War data set~\cite{sarkees:wayman:2010} (Fig.~\ref{fig:cow:data}). This data set provides comprehensive coverage in this period, with few artifacts and relatively low measurement bias, and allows us to focus on a period during which Richardson's model is plausible~\cite{cederman:etal:2011}.

\section{Preliminaries}

Before analyzing any data, we must clarify several epistemological issues and the impact of different assumptions on the accuracy and interpretation of the analysis.

Power-law distributions have unusual mathematical properties~\cite{newman:2005}, which can require specialized statistical tools to analyze. (For a primer on power-law distributions in conflict, see Refs.~\cite{cederman:2003,clauset:etal:2007}.) For instance, when observations are generated by a power law, time series of summary statistics like the mean or variance can exhibit long fluctuations resembling a trend. The largest and longest fluctuations occur for a scaling parameter of $\alpha < 3$, when one or both the mean and variance are mathematically infinite, i.e., they never converge, even for infinite-sized samples. For interstate wars, this property could produce long transient patterns of low-severity or the absence of wars, making it difficult to distinguish a genuine trend toward peace from a mere fluctuation in a stationary process.

To illustrate the counter-intuitive nature of power-law distributions, consider a world where the heights of Americans are power-law distributed, but with the same average as reality (about 1.7 meters), and we line them up in a random order. In this world, nearly 60,000 Americans would be as tall as the tallest adult male on record (2.72 meters), 10,000 individuals would be as tall as an adult male giraffe, one would be as tall as the Empire State Building (381 meters), and 180 million diminutive individuals would stand only 17 cm tall. As we run down the line of people, we would repeatedly observe long runs of relatively short heights, one after another, and then, rarely, we would encounter a person so astoundingly tall that their singular presence would dramatically shift our estimate of the average or variance of all heights. This is the kind of pattern that we see in the sizes of wars.

Identifying a power-law distribution within an empirical quantity can suggest the presence of exotic underlying mechanisms, including nonlinearities, feedback loops, and network effects~\cite{newman:2005}, although not always~\cite{reed:hughes:2002}, and power laws are believed to occur broadly in complex social, technological, and biological systems~\cite{clauset:etal:2009}. For instance, the intensities or sizes of many natural disasters, such as earthquakes, forest fires, and floods~\cite{newman:2005}, as well as many social disasters, like riots and terrorist attacks~\cite{biggs:2005,clauset:etal:2007}, are well-described by power laws.

Testing if some quantity does or does not follow a power law requires specialized statistical tools~\cite{resnick:2006,clauset:etal:2009}, because uncertainty tends to be greatest in the upper tail, which governs the frequency of the largest and rarest events, i.e., the frequency of large wars. Modern statistical tools provide rigorous methods for estimating and testing power-law models, distinguishing them from other ``heavy-tailed'' distributions, and even using them to make statistical forecasts of future events~\cite{clauset:woodard:2013}.

Poisson processes pose fewer statistical issues than power-law distributions. However, for consistency, our analysis applies similar methods to both war timing and size data. Specifically, we estimate an ensemble of models, each fitted to a bootstrap of the empirical data, which better represents our statistical uncertainty than would a single model. Technical details are described in Ref.~\cite{clauset:2018}.

Richardson's models are defined in terms of absolute numbers, i.e., the number of interstate wars per year and the number of battle deaths per war. Hence, we consider war variables in their unnormalized forms and consider all recorded interstate wars, meaning that our analysis takes the entire world as a system.

Analyses of interstate war statistics sometimes normalize either the number of wars or their sizes by some kind of reference population, e.g., dividing a war's size by the global population at the time. Such normalizations represent additional theoretical assumptions about the underlying data generating process.

For instance, normalizing the number of wars per year by the number of pairs of nations that could be at war assumes that war is a dyadic event and that dyads independently generate conflicts with equal probability~\cite{ward:etal:2007}. This choice of normalizer grows quadratically with the number of nations, and will create the appearance of a trend toward fewer wars, even if Richardson's stationary model of wars is correct in an absolute sense. Considerable evidence indicates that dyads do not independently generate conflicts, and that dyadic likelihoods vary across time and space, and by national covariates~\cite{ray:1998,ward:etal:2007,leeds:2003,alesina:spolaore:1997,jackson:nei:2015}.

Similarly, normalizing a war's size by the nations' or world's total population, producing a per-capita rate~\cite{pinker:2011,cirillo:taleb:2016}, assumes that individuals contribute independently and with equal probability to potential or actual violence, regardless of who or where they are. Normalizing by world population is thus equivalent to assuming that doubling Canada's population would linearly increase the level of violence in a war in Yemen. In general, human populations have increased so dramatically over the past 200 years that this normalizer nearly always produces the appearance of a decline in violence, even if wars were, in an absolute sense, stationary. However, there is little evidence that real conflict sizes or rates increase linearly with population~\cite{bowles:2009,oka:etal:2017,falk:hildebolt:2017}.

That said, per-capita variables can be useful for other reasons~\cite{pinker:2011}, and population surely does play some role in the sizes of wars, albeit probably not a simple one~\cite{oka:etal:2017,falk:hildebolt:2017}. A realistic per-capita normalizer should instead account for the effects of alliances, geographic proximity, geopolitical stability, democratic governance, economic ties, etc.~\cite{cederman:etal:2011}, in addition to population, and would be akin to modeling the underlying processes that generate interstate wars. This represents an important direction for future work.

\section{The Size and Timing of Wars}

\begin{figure}[b!]
\begin{center}
\includegraphics[scale=0.455]{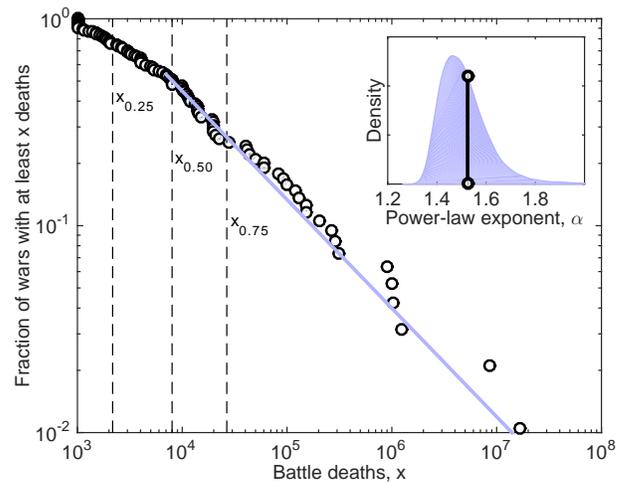}
\end{center}
\vspace{-3mm}
\caption{\textbf{Interstate wars sizes, 1823--2003.} The maximum likelihood power-law model of the largest-severity wars (solid line, $\hat{\alpha}=1.53\pm0.07$ for $x\geq \hat{x}_{\min}=7061$) is a statistically plausible data-generating process of these 51 empirical severities (Monte Carlo, $p_{\rm KS}=0.78\pm0.03$). For reference, distribution quartiles are marked by vertical dashed lines.  Inset:\ bootstrap distribution of maximum likelihood parameters $\Pr(\hat{\alpha})$, with the empirical value (black line). Originally published in Ref.~\cite{clauset:2018}.
\label{fig:war:sizes}  } 
\end{figure}

In Richardson's view, the size or severity of an interstate war (number of battle deaths) follows a power-law distribution of the form $\Pr(x)\propto x^{-\alpha}$ for some $\alpha>1$ and for all $x\geq x_{\min} > 0$.

Using standard techniques to estimate $\alpha$ and $x_{\min}$, and to test the fitted distribution~\cite{clauset:etal:2009}, we find that the set of observed interstate wars sizes, from 1823--2003, are statistically indistinguishable from an iid draw from a power-law distribution (Fig.~\ref{fig:war:sizes})~\cite{clauset:2018}. 

Similarly, Richardson posits that the onset of a new interstate war follows a Poisson process, meaning that wars arrive at a constant rate $q$, and the time $t$ between onsets of new wars follows a geometric distribution of the form $\Pr(t)\propto \textrm{e}^{-q t}$ for some $q>0$ and for all $t \geq 1$.

\begin{figure}[t!]
\begin{center}
\includegraphics[scale=0.455]{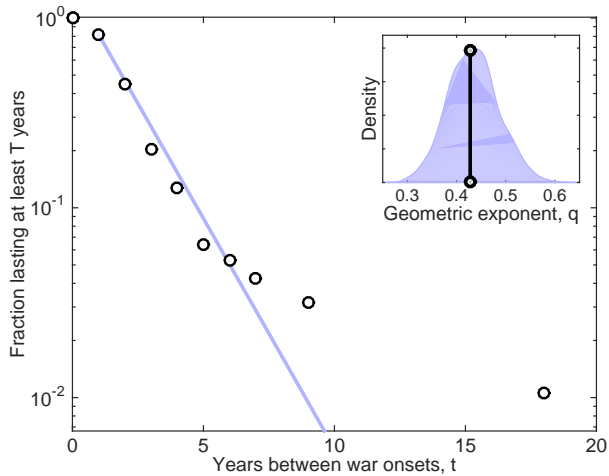}
\end{center}
\vspace{-3mm}
\caption{\textbf{Times between interstate war onsets, 1823--2003.} The maximum likelihood geometric model (solid line, $\hat{q}=0.428\pm0.002$ for $t\geq 1$) is a statistically plausible data-generating process of the empirical delays (Monte Carlo, $p_{\rm KS}=0.13\pm0.01$), implying that the apparent discontinuity at $t=5$ is a statistical artifact. Inset:\ the bootstrap distribution of maximum likelihood parameters $\Pr(\hat{q})$, with the empirical estimate (black line). Originally published in Ref.~\cite{clauset:2018}.
\label{fig:peace:pdf} } 
\end{figure}

Using the same techniques as above to estimate $q$ and test the fitted distribution, we find that the set of observed delays between onsets are statistically indistinguishable from an iid draw from a Poisson process (Fig.~\ref{fig:peace:pdf})~\cite{clauset:2018}. 

That is, both the size and timing of interstate wars 1823--2003 are statistically indistinguishable from Richardson's simple model of a power-law distribution for war sizes and a Poisson process for their arrival~\cite{richardson:1944,richardson:1960}. This agreement is remarkable considering the overall simplicity of the model compared to the complexity and contingency of international relations over this period, and the fact that this time period includes nearly 60 years of additional data over Richardson's original analysis.

\section{Are Large Wars Declining?}

Combining the two parts of Richardson's model allows us to generate simulated interstate war data sets, drawn from a stationary process, with similar onset times and war sizes as the historical record. The statistics of these simulated histories define a reference distribution against which we can compare aspects of the historical record.

We now apply this model to address a long-running debate in international relations:\, did the underlying processes that generate interstate wars change after the Second World War? This point in history is commonly proposed as the beginning of a ``long peace'' pattern in interstate conflict, meaning a pronounced decrease in the frequency and severity of wars, especially large ones~\cite{gaddis:1986,ward:etal:2007,levy:thompson:2010,pinker:2011,braumoeller:2013}.

To test the long peace hypothesis from Richardson's perspective, we consider the accumulation of large interstate wars over time, and assess whether or when that accumulation in the post-war period represents a low probability event under the stationary model. If the accumulation during the long peace is statistically unusual under the reference distribution, it would indicate support for an underlying change in the generating processes.

\begin{figure}[t]
\begin{center}
\includegraphics[scale=0.455]{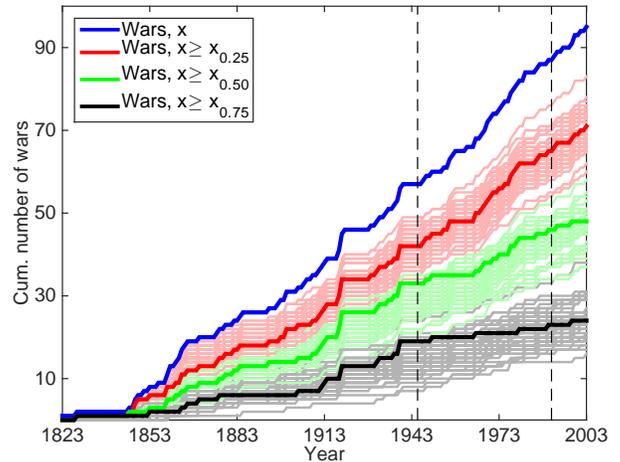} 
\end{center}
\vspace{-4mm}
\caption{\textbf{Historical and simulated accumulation curves of interstate wars.} Empirical cumulative counts of wars of different sizes (dark lines) over time, alongside ensembles of simulated counts from a stationary model (light lines), in which empirical severities are replaced iid with a bootstrap draw from the empirical severity distribution. Dashed lines mark the end of the Second World War and the end of the Cold War. Originally published in Ref.~\cite{clauset:2018}. \label{fig:war:overTime:2} } 
\end{figure}

We define ``large'' wars as those in the upper quartile of the historical war size distribution, meaning $x\geq x_{0.75}=26,625$ battle deaths, but similar results are obtained for other large thresholds. The initial 1823--1939 period contains 19 such large wars, for an arrival rate of one per 6.2 years, on average. The ``great violence'' pattern of 1914--1939, which spans the onsets of the First and Second World Wars, includes 10 large wars, or about one every 2.7 years. The long peace of the 1940--2003 post-war period contains only 5 large wars, or about one every 12.8 years. Figure~\ref{fig:war:overTime:2} shows the historical accumulation curves of these events, and for smaller wars, as a function of time.

Our combined model takes the historical war onset years as given, and then for each of these 95 conflicts, draws a synthetic war size iid from the empirical size distribution (with replacement), as in a simple bootstrap of the data. Ref.~\cite{clauset:2018} considers two additional models of this flavor, which produce larger variances in the accumulation curves for large wars but yield similar results and conclusions.

\begin{table*}[t!]
\begin{tabular}{ll|c}
Empirical pattern & Formalization & Model \\ 
\hline
great violence & $\Pr(V\equiv n\geq10$ large wars over $t\leq 27$ years)~ & ~$0.107(1)$\hspace{0.5em}   \\
long peace & $\Pr(P\equiv n \leq5$ large wars over $t\geq 64$ years) & ~$0.622(2)$\hspace{0.5em}
\end{tabular}
\caption{\textbf{Stationary likelihood of empirical conflict patterns.} Under a simple stationary model of conflict generation (see text), the estimated likelihoods of observing two particular large-war patterns over the period 1823--2003:\ a ``great violence,'' meaning 10 or more large war onsets ($x\geq x_{0.75}$) over a 27 year period (the empirical count of such onsets, 1914--1939); or, a ``long peace,'' meaning 5 or fewer large war onsets over a 64 year period (the empirical count of such onsets, 1940--2003). Probabilities estimated by Monte Carlo. Parenthetical values indicate the standard error of the least significant digit. Originally published in Ref.~\cite{clauset:2018}. \label{table:probs}}
\end{table*}

\subsection{Evaluating the past}

Within the historical accumulation curve for large wars, the long peace is a visible pattern, in which the arrival rate (the curve's slope) is substantially more flat than in the preceding great violence period (Fig.~\ref{fig:war:overTime:2}). However, under the stationary model, this pattern is well within the envelope of simulated curves, and the observed pattern is statistically indistinguishable from a typical excursion, given the heavy-tailed nature of historical war sizes.

In fact, most simulated war sequences contain a period of ``peace'' at least as long in years and at least a peaceful in large-war counts as the long peace (Table~\ref{table:probs}). Fifty years or more of relatively few large wars is thus entirely typical, given the empirical distribution of war sizes, and observing a long period of peace is not necessarily evidence of a changing likelihood for large wars~\cite{cirillo:taleb:2016,clauset:2018}. Even periods comparable to the great violence of the World Wars are not statistically rare under Richardson's model (Table~\ref{table:probs}).

Taking these findings at face value implies that the probability of a very large war is constant. Under the model, the 100-year probability of at least one war with $16,634,907$ or more battle deaths (the size of the Second World War) is $0.43\pm0.01$, implying about one such war per 161 years, on average.

\subsection{Peering into the future}

We can also use Richardson's model to simulate future sequences of interstate wars, and thereby evaluate how long the long peace must last before it becomes compelling evidence that the underlying processes did, indeed, change after the Second World War. To extend the simulated war sequences beyond 2003, for each year, we create a new war onset according to a simple Bernoulli process, with the historical production rate (on average, a new war every 1.91 years). We then linearly extrapolate the long peace pattern, in which a new large war occurs on average every 12.8 years, out into the future until 95\% of the simulated accumulation curves exceed the extrapolated pattern's curve. At that moment in time, the long peace pattern will have become statistically significant, by conventional standards, relative to the stationary model, and we could say with confidence that the time since the Second World War was governed by a different, more peaceful underlying process.

\begin{figure}[t]
\begin{center}
\includegraphics[scale=0.455]{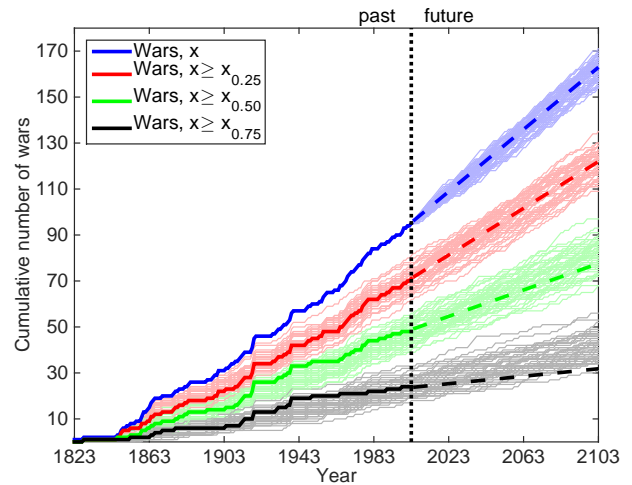} 
\end{center}
\vspace{-4mm}
\caption{\textbf{How long must the peace last?} Simulated accumulation curves for wars of different sizes under a simple stationary model, overlaid by the empirical curves up to 2003 (dark lines) and linear extrapolations of the empirical post-war trends (the long peace) for the next 100 years (dashed lines). Originally published in Ref.~\cite{clauset:2018}. \label{fig:war:overTime:3} } 
\end{figure}

In this extrapolated future, the post-war pattern of relatively few large wars becomes progressively more unlikely under a stationary hypothesis (Fig.~\ref{fig:war:overTime:3}). However, it is not until 100 years into the future that the long peace becomes statistically distinguishable from a large but random fluctuation in an otherwise stationary process. Even if there were no large wars anywhere in the world after 2003, the year of significance would arrive only a few decades earlier.

The consistency of the historical record of interstate wars with Richardson's stationary model places an implicit upper bound on the magnitude of change in the underlying conflict generating process since the end of the Second World War~\cite{cederman:2001}. Our modeling effort here cannot rule out the existence of a change in the rules that generate interstate conflicts, but if it occurred, it cannot have been a dramatic shift. The results here are entirely consistent with other evidence of genuine changes in the international system, but they constrain the extent to which such changes could have genuinely impacted the global production of interstate wars.

\section{Discussion}

The agreement between the historical record of interstate wars and Richardson's simple model of their frequency and severity is truly remarkable, and it stands as a testament to Richardson's lasting contribution to the study of violent political conflict.

There are, however, a number of caveats, insights, and questions that come out of our analysis. For instance, Richardson's Law---a power-law distribution in conflict event sizes---appears to hold only for sufficiently large ``deadly quarrels,'' specifically those with $7061$ or more battle deaths. The lower portion of the distribution is slightly more curved than expected for a simple power law, which suggests potential differences in the processes that generate wars above and below this threshold.

With only 95 conflicts and a heavy-tailed distribution of war sizes, there are relatively few large wars to consider. This modest sample size surely lowers the statistical power of any test and is likely partly to blame for needing nearly 100 more years to know whether the long peace pattern is more than a run of good luck under a stationary process.

One could imagine increasing the sample size by including civil wars, which are about three times more numerous than interstate wars over 1823--2003. Including these, however, would confound the resulting interpretation, because civil wars have different underlying causes~\cite{salehyan:gleditsch:2006,cederman:etal:2013,wucherpfennig:etal:2016}, and because the distribution of civil war sizes is shifted toward smaller conflicts and exhibits relatively fewer large ones~\cite{lacina:2006}.

Putting aside these technical issues, the larger question our analysis presents is this:\ how can it be possible that the frequency and severity of interstate wars are so consistent with a stationary model, despite the enormous changes and obviously non-stationary dynamics in human population, in the number of recognized states, in commerce, communication, public health, and technology, and even in the modes of war itself? The fact that the absolute number and sizes of wars are plausibly stable in the face of these changes is a profound mystery for which we have no explanation.

There is, of course, substantial evidence for a genuine post-war trend toward peace, based on mechanisms that reduce the likelihood of war~\cite{ray:1998,leeds:2003,jackson:nei:2015} and on statistical signatures of a broad and centuries-long decline in general violence~\cite{gurr:2000,payne:2004,goldstein:2011,pinker:2011} or the improvement of other aspects of human welfare~\cite{roser:etal:2017}.

But, a full accounting of the likelihood that the long peace will endure must also consider mechanisms that increase the likelihood of war (for example, Refs.~\cite{bremmer:1992,mansfield:snyder:1995,barbieri:1996}). War-promoting mechanisms certainly include the reverse of established peace-promoting mechanisms, e.g., the unraveling of alliances, the slide of democracies into autocracy, and the fraying of economic ties, but they may also include unknown mechanisms.

In the long run, processes that promote interstate war may be consequences of those that reduce it over the shorter term, through feedback loops, tradeoffs, or backlash effects. For example, the persistent appeal of nationalism, whose spread can increase the risk of interstate war~\cite{schrockjacobson:2012}, is not independent of deepening economic ties via globalization~\cite{smith:1992}. Investigating such interactions is a vital direction of future research, and will facilitate a more complete understanding of the processes that govern the likelihood of patterns like the long peace.

More concretely, our results here indicate that the post-war efforts to reduce the likelihood of large interstate wars have not yet changed the observed statistics enough to tell if they are working. This fact does not lessen the face-value achievement of the long peace, as a large war today between major powers could be very large indeed, and there are real benefits beyond lives saved~\cite{roser:etal:2017} that have come from increased economic ties, peace-time alliances, and the spread of democracy. However, it does highlight the continued relevance of Richardson's foundational ideas as the appropriate null hypothesis for patterns in interstate war.

One explanation for the apparent stationarity of wars since 1823 is the existence of compensatory trends in related conflict variables that mask a genuine change in the conflict generating processes. Patterns across multiple conflict variables do seem to indicate a broad shift toward less violence~\cite{gurr:2000,payne:2004,goldstein:2011,pinker:2011}. But, not all conflict variables support this conclusion, and some, such as military disputes and the frequency of terrorism, seem to be increasing, instead~\cite{harrison:wolf:2012,clauset:woodard:2013}. Untangling conflict variables' interactions and characterizing their trends and differences across groups of nations will be a valuable line of future work.

An alternative explanation is that the mechanisms that govern the likelihood of war have unfolded heterogeneously across time and geographic regions over the past 200 years, thereby creating an illusion of global stationary by coincidence. The long peace pattern is sometimes described only in terms of peace among largely European powers, who fell into a peaceful configuration after the great violence for well understood reasons. In parallel, however, conflicts in other parts of the world, most notably Africa, the Middle East, and Southeast Asia, have became more common, and these may have statistically balanced the books globally against the decrease in frequency in the West, and may even be causally dependent on the drivers of European war and then peace.

Developing a more mechanistic understanding of how changes in the likelihood of conflict in one part of the world  may induce compensatory changes in the likelihood of conflict in other parts of the world would have enormous value if it could explain how some regions can fall into more peaceful patterns as a group, while other regions go the opposite direction. The evident stability of Richardson's Law may, in the end, be an artifact of these kinds of complex, macro-scale dynamics, playing out across the global stage.

Finally, it is worth reiterating how remarkable and counter-intuitive it is that Richardson's original models of the frequency and severity of wars, first proposed more than half a century ago, successfully hold up under modern, more rigorous statistical methods of evaluation applied to far more comprehensive data. A sobering implication of this success is that the probability of a large interstate appears to have remained constant, despite profound collective efforts to lower it.

More importantly for the general study of conflict, Richardson's work presents a simple and enduring mystery: how can the frequency and severity of interstate wars be so consistent with a stationary model, despite the dramatic changes and obviously non-stationary dynamics in so many other aspects of human civilization? Answering this question will shed new light on the underlying causes of war, and greatly inform efforts to promote peace. Richardson, who was an avowed pacifist and who worked as an ambulance driver during the First World War, would surely be pleased if his work on the statistics of war ultimately helped devise better policies to promote peace.

\end{document}